\documentclass[longbibliography,
reprint,
superscriptaddress,
amsmath,amssymb,
aps,
prx, floatfix
]{revtex4-1}
\usepackage{amsmath, graphicx,multirow, verbatim, psfrag}
\usepackage{amssymb, amsthm}
\usepackage{floatpag}
\usepackage[margin=18mm]{geometry}
\usepackage{fancyhdr}
\usepackage{wrapfig}
\fancypagestyle{newstyle}{
\fancyhf{} 
\fancyfoot[R]{\thepage} 

}
\theoremstyle{remark}

\makeatletter
\@ifundefined{textcolor}{}
{%
 \definecolor{BLACK}{gray}{0}
 \definecolor{WHITE}{gray}{1}
 \definecolor{RED}{rgb}{1,0,0}
 \definecolor{GREEN}{rgb}{0,0.5,0}
 \definecolor{BLUE}{rgb}{0,0,1}
 \definecolor{CYAN}{cmyk}{1,0,0,0}
 \definecolor{MAGENTA}{cmyk}{0,1,0,0}
 \definecolor{YELLOW}{cmyk}{0,0,1,0}
 }
  \usepackage[usenames,dvipsnames]{xcolor} 

\usepackage{subfig,subfloat,float}
  
\begin{document}
\title{Unifying Theories in High-Dimensional Biology: Approaches, Challenges and Opportunities}
\author{Marianne Bauer}
\affiliation{
Bionanoscience Department, Kavli Institute of Nanoscience, Delft University of Technology, van der Maasweg 9, 2629 Delft, the Netherlands}
\email{m.s.bauer@tudelft.nl}
\author{Akshit Goyal}
\affiliation{International Centre for Theoretical Sciences, Tata Institute of Fundamental Research, Bengaluru 560089, India.}
\author{Sidhartha Goyal}
\affiliation{Department of Physics, University of Toronto, Toronto, Ontario, Canada}
\author{Gautam Reddy}
\affiliation{Joseph Henry Laboratories of Physics, Princeton University,
Princeton, New Jersey 08544, USA}
\author{Shaon Chakrabarti}
\affiliation{National Centre for Biological Sciences, Tata Institute of Fundamental Research, Bengaluru 560065, India.}
\author{Michael M Desai}
\affiliation{Departments of Organismic and Evolutionary Biology and of Physics, Harvard University, Cambridge MA 02138, USA}
\author{William Gilpin}
\affiliation{Department of Physics, University of Texas at Austin, Austin, Texas, USA}
\author{Jacopo Grilli}
\affiliation{The Abdus Salam International Centre for Theoretical Physics (ICTP), Strada Costiera 11, 34014 Trieste, Italy}
\author{Kabir Husain}
\affiliation{Department of Physics and Astronomy, and Laboratory for Molecular Cell Biology, University College London, London UK}
\author{Sanjay Jain}
\affiliation{Department of Physics and Astrophysics, University of Delhi, Delhi 110007 India and Santa Fe Institute, 1399 Hyde Park Road, Santa Fe, NM 87505, USA}
\author{Mohit Kumar Jolly}
\affiliation{Department of Bioengineering, Indian Institute of Science, Bengaluru 560012, India}
\author{Kyogo Kawaguchi}
\affiliation{Nonequilibrium Physics of Living Matter Laboratory, Pioneering Research Institute, RIKEN, Kobe, 650-0047, Japan}
\affiliation{Institute for Physics of Intelligence, Department of Physics, The University of Tokyo, Tokyo, 113-0033, Japan}
\author{Aneta Koseska}
\affiliation{Cellular Computations and Learning, Max Planck Institute for Neurobiology of Behavior-caesar, Bonn, Germany}
\author{Milo Lin}
\affiliation{Lyda Hill Department of Bioinformatics, University of Texas Southwestern Medical Center, Dallas, TX, USA}
\author{Leelavati Narlikar}
\affiliation{Departments of Data Science and Biology, Indian Institute of Science Education and Research, Pune 411008, India}
\author{Simone Pigolotti}
\affiliation{Biological Complexity Unit, Okinawa Institute of Science and Technology, Onna, Okinawa 904-0495, Japan.}
\author{Archishman Raju}
\affiliation{National Centre for Biological Sciences, Tata Institute of Fundamental Research, Bengaluru 560065, India.}
\author{Krishna Shrinivas}
\affiliation{Department of Chemical and Biological Engineering, Northwestern University, Evanston IL USA}
\author{Rahul Siddharthan}
\affiliation{The Institute of Mathematical Sciences, Chennai 600113, India and Homi Bhabha National Institute, Mumbai 400094, India}
\author{Greg J. Stephens}
\affiliation{Biological Physics Theory Unit, OIST Graduate University, Okinawa 904-0495, Japan}
\affiliation{Department of Physics and Astronomy, Vrije Universiteit Amsterdam, 1081HV Amsterdam, The Netherlands}
\author{Andreas Tiffeau-Mayer}
\affiliation{Division of Infection and Immunity \& Institute for the Physics of Living Systems, University College London, London WC1E 6BT, UK}
\author{Suriyanarayanan Vaikuntanathan}
\affiliation{The James Franck Institute, The University of Chicago, Chicago,
IL, USA}
\affiliation{Department of Chemistry, The University of Chicago, Chicago, IL, USA}

\begin{abstract}
Across biological subdisciplines, the last decade has seen an explosion of high-dimensional datasets, including datasets for cells, species, immune systems, neurons and behaviour. At the ICTS workshop `Unifying Theories in High-Dimensional Biophysics', we discussed whether this high dimensionality poses a challenge or opportunity for describing, understanding and predicting biological systems theoretically. We discussed methods, models and frameworks that can help with addressing empirical observations based on these high-dimensional datasets. In this Comment, we summarize the challenges and opportunities that emerged in discussions according to individual participants.
\end{abstract}

\maketitle

\clearpage
\section{Introduction}
The high-dimensional nature of biological systems creates a formidable challenge in making conceptual advances. The vast amount of datasets across diverse fields, ranging from development, immunology, ecology, evolution, neuroscience and behavior \cite{Lieberman-Aiden,tang2009mrna,treutlein2014reconstructing,lubeck2012single,sureshchandra2024tissue,thompson2017communal,sunagawa2020tara,human2012structure,international2025brain, mckenzie2025capturing} offer an opportunity to glean quantitative insights.  Nevertheless, we lack common theoretical frameworks to extract generalizable principles.

One central idea emerging from recent work is that low-dimensional models can be surprisingly good at explaining empirical observations while providing specific testable predictions. Examples include the use of sloppy models to create simplifying descriptions of biochemical networks \cite{gutenkunst,transtrum2011geometry, transtrum2014model}, coarse-grained resource allocation models in microbial metabolism \cite{scott2010interdependence,scott2023shaping}, low-dimensional dynamical systems models of development \cite{corson2017gene, rand2021geometry}, or low-dimensional structure in genotype-phenotype maps \cite{petti2023inferring, johnson2023epistasis}, amongst others.  Machine learning helps finding some of this low-dimensional structure, for example, by using sequence data from developing organisms \cite{van2020single}, tumors and immune repertoires \cite{pyo2025data}, and from large-scale neural recordings \cite{meshulam2025statistical, zhong2025unsupervised}.  However, different subfields have developed different analytical and machine learning approaches for tackling high-dimensional complexity. 

In this brief interdisciplinary viewpoint, we share thoughts discussed in the realm of high-dimensional biophysics at a workshop at the International Centre for Theoretical Sciences (ICTS) in the summer of 2025.
Is there a common mathematical language that can be used to describe emergent simplicity across systems? Are there common biological principles, such as evolvability and functional robustness, that make low-dimensional descriptions possible? How do we understand if a system truly operates in high-dimensions, for example, the fine-scale diversity in microbial ecosystems? What are good ‘null’ models, and what details matter for explaining phenomenology? 

We gathered contributions from theoretical and computational physicists at our workshop working in diverse areas of biology, including development, ecology and evolution, immunology, neuroscience, protein science. These contributions  highlight commonalities, differences, and challenges of high dimensional biology and identify opportunities for progress towards a unifying framework. We structured individual contributions by scale of their dominant application topics. 

\section{Molecules and Proteins}

\subsection{Kabir Husain}

A physicist once interrupted a talk I was giving on protein chaperones with a question—asked with a certain degree of incredulity. How, they wanted to know, could a single chaperone help so many different proteins fold? Did it somehow memorize the correct folded configuration of each one?

Indeed, recognizing that biology operates in what we might call a “high-dimensional” space raises unsettling questions of this kind. One defining feature of a high-dimensional space is how easy it is to get lost within it: while random diffusion might suffice to find a target in one or two dimensions, the likelihood drops precipitously as the number of dimensions increases. This would seem to suggest that robust, reliable systems should be low-dimensional, with only a few moving parts. Yet biology appears to defy this intuition. It is full of diverse, intricately structured entities that assemble precisely and reproducibly—whether it be the sequence of nucleotides in a genome, the arrangement of atoms in a folded protein, or the patterning of cells in a developing organism. How does that work?

Questions of this breadth inevitably have many system-specific answers. But some overarching strategies can be discerned. First, the “correct” or functional biological configuration often corresponds to one that is energetically most favourable—as in protein folding—or that has the longest residence time, as in immune recognition. Second, biological search processes are rarely random walks. They follow stereotyped trajectories through energy landscapes, guided by structures such as folding funnels or Waddington’s developmental “landscape.” Third—and perhaps least generalizable—when errors occur, biology often has ways of detecting them without needing to know exactly what went wrong. For instance, some chaperones recognize exposed hydrophobic patches that would normally be buried within a protein’s core, while DNA repair mechanisms detect distortions in the regular shape of Watson–Crick base pairs. In short, the physical architecture of biological systems seems remarkably well adapted to navigating high-dimensional spaces reliably.

What do we gain by thinking at this level of generality? Beyond the intellectual pleasure of tracing analogies across systems, such reflections suggest constraints on what kinds of physical systems can support something we would recognize as “biology.” If evolution—another search through an immense, high-dimensional space—were to begin again on a different world, many details would surely differ. However, I suspect that the basic challenge of finding one’s way through high-dimensional spaces would impose severe constraints on the physical make-up of any living system.

\subsection{Milo Lin}
Proteins are chains of hundreds to thousands of amino acids that self-fold into their functional configurations. Because proteins carry out most of the biomolecular processes responsible for life, how the amino acid sequence encodes function is the high-dimensional question upon which many other questions, at larger scales, are built upon. Yet this question is largely unresolved for two reasons. First, although much progress has been made in predicting protein structure from sequence, these advances are black boxes that have not yielded unifying principles of protein folding. Second, the assumption that structure (i.e. protein folding) is sufficient to prescribe function is incomplete. For enzymatic proteins, the activity of the active site is often coupled to the dynamics of the rest of the protein. Many proteins are allosteric, which means that they can modulate the activity of their active site depending on detection of signals at other sites of the protein. This allows proteins to be input-output devices, which is crucial for their role as the building blocks of biological circuits. In many proteins, allostery depends on correlated dynamics. We have found that treating proteins as coupled networks of conformational fluctuations enables quantitative prediction of protein allostery. At the level of amino acids, protein fluctuations can be well-approximated as digital transformations between discrete states, allowing us to borrow approaches from spin systems and glasses. This meeting introduced connections of protein functional with other fields, in particular ecological networks, and I anticipate future progress to be advanced by cross-application of ideas and approaches between these fields operating at very different scales.

\subsection{Suriyanarayanan Vaikuntanathan}
High-dimensionality poses a fundamental challenge in deciphering complex biological codes like the glycan code, where the combinatorial diversity of glycan structures creates a potentially overwhelming information space that cells must somehow read and interpret. Our recent work in Nature Communications \cite{floyd2025limits} demonstrates how principles from non-equilibrium thermodynamics offer a promising way forward: molecular promiscuity in receptor-glycan interactions can actually serve as a feature rather than a bug, effectively reducing the dimensionality required for accurate biological readout. This insight connects naturally to techniques from computer science, particularly compressed sensing, which shows how high-dimensional signals can be recovered from surprisingly few measurements when the underlying structure is sparse or compressible. More broadly, I see potential for bidirectional exchange between fields—while computational frameworks like compressed sensing illuminate how biological systems might efficiently decode complex molecular information, the thermodynamic principles governing promiscuous binding and selective readout in biology could inspire novel computational paradigms that go beyond conventional compressed sensing approaches. These bio-inspired algorithms might offer new ways to handle high-dimensional data in artificial systems, particularly where noise, dynamics, and energy constraints mirror those faced by living cells.

\section{The nucleus: chromatin and gene expression}
\subsection{Kyogo Kawaguchi}
Over the past decades, biology has entered a high-dimensional era, with massive data collection across gene expression dynamics, chromatin profiles, neuronal activity, and more. This expansion partly realizes the vision of nonlinear dynamics and self-organization theory, which converged into systems biology around the 2000s. With this abundance of measurable quantities, a central challenge now is to ask whether we can already constrain which models are sensible and which are falsifiable. 

Since biological dynamics are inevitably multiscale and complex, it has been natural to begin from a common framework such as dynamical-systems approaches derived from chemical reaction networks, which have been particularly influential in explaining cell fate decisions. But these models often neglect spatial components—such as phase separation in the nucleus and polymer dynamics in chromatin—that almost certainly contribute to fate determination. Other strategies from physics, such as coarse-graining and the use of phase transitions and macroscopic order parameters, have proven useful for capturing at least parts of those phenomena. Beyond these, heuristic and data-driven approaches are rapidly advancing, highlighting the need for still broader theoretical frameworks that can connect diverse mechanistic models with purely empirical descriptions.

\subsection{Leelavati Narlikar and Rahul Siddharthan}
Modern molecular biology is driven by high-throughput data generation, starting with the advent of new "next-generation" sequencing technologies about 20 years ago. These technologies, apart from sequencing genomes, can be used to perform genome-wide assays for transcript abundance, DNA-protein binding, chromatin accessibility, chromatin interactions, and much more. More recently, single-cell transcriptomics has enabled generation of transcriptomic and chromatin accessibility data at a single-cell level.  The output of such experiments consists of matrices with 1000-100,000 rows (cells) and up to 25,000 columns (genes). Analysis of such inherently high-dimensional data, and interpretation of the results, is a challenge, and the field relies on methods such as UMAP for visualization  which have caused some controversy \cite{chari2023specious,grobecker2024identifying}. 

Our talks focused on the specific problems of DNA-protein interaction and transcription initiation, where current datasets are mostly from bulk samples. Even so, genome-wide data exists for 100+ transcription actors and histone modifications, and dozens of cell types, in human alone. Predictive models for DNA-protein binding, including sequence information, interactions among TFs, chromatin accessibility data, have been a challenge for many years. We discussed combinatorial binding via mixture models, and sequence signatures identified via deep learning. In the near future, single-cell ChIP-seq is expected to become more common, enabling us to learn much more specific modes of binding but also increasing the challenge of analysing the data.  

At this meeting, we heard from other speakers about the application of dimensionality reduction, deep learning methods, Bayesian methods, information theory, and more, across multiple fields. In particular, the use of latent dimensions to compress data and better understand it was discussed in multiple talks. This is something we also spoke about in our mixture models of regulatory data. We expect that such techniques will be developed synchronously across many areas of modern biology, with ample cross-pollination of ideas.

\begin{figure}
\begin{center}
\includegraphics[width=\linewidth]{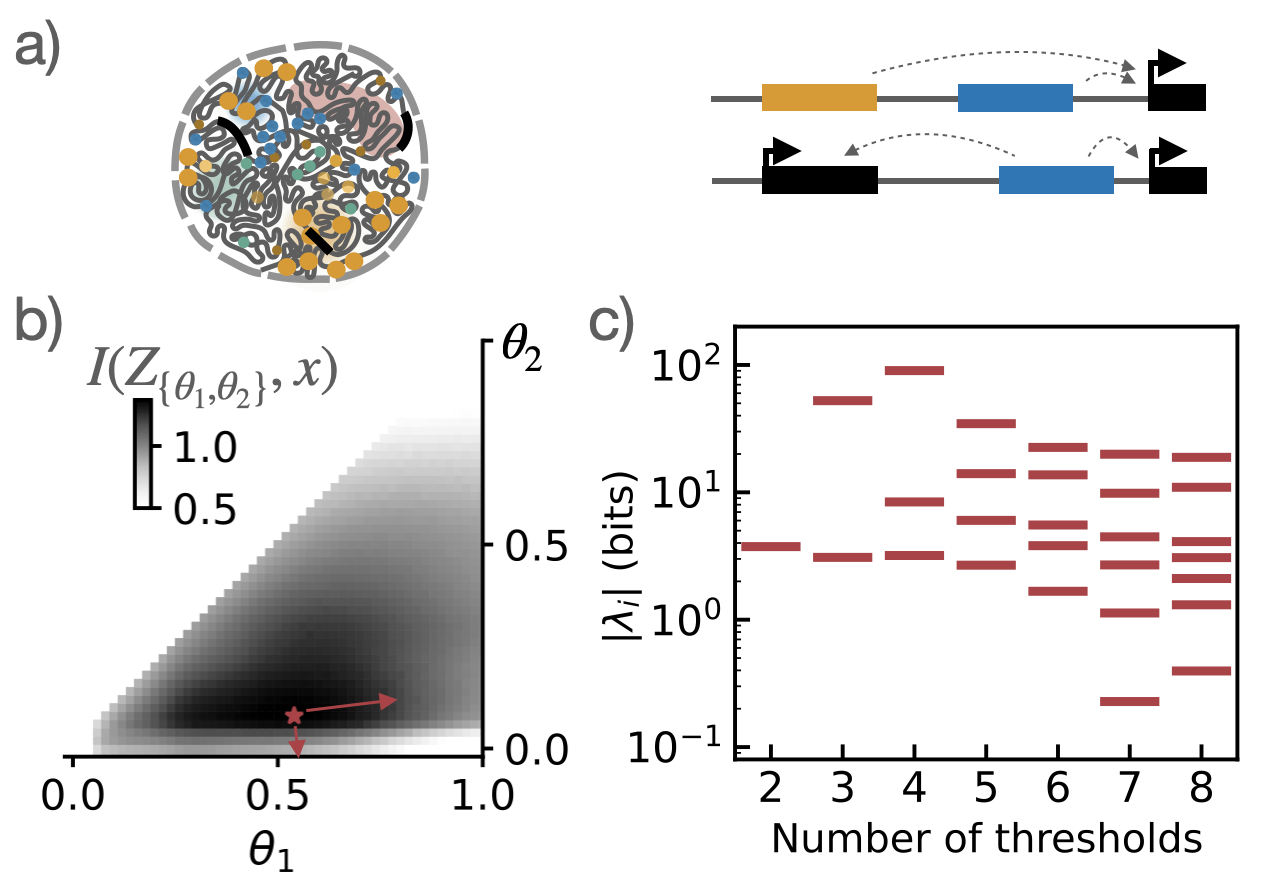}
\caption{{\bf Sloppy landscapes of optimal gene regulation.} a) Genes are expressed in the spatially complex geometry of the nucleus, based on multiple genomic regions upstream and/or downstream of the gene that can regulate a single gene together as well as multiple genes. b) Even in the simplest possible model where these regulatory regions activate expression based on threshold concentrations $\theta$ of an input, the optimal positioning of these thresholds that maximizes transfer of developmental information $I(Z_{\{\theta_1,\theta_2\}};x)$ is `sloppy’: c) This implies that the eigenvalues $\lambda$ of the Hessian matrix at the optimum span several decades (adapted from Ref \cite{Baueretal}).
}
\label{fig:generegulation}
\end{center}
\end{figure}

\subsection{Marianne Bauer}
Genes are regulated by a myriad of processes, involving the binding of transcription or pioneer factors to one or multiple enhancer or promoter regions that may require spatial contacts or the establishment of a local transcription factor hub \cite{furlonglevine_18} (see Fig.\ref{fig:generegulation}). This regulation is necessary for biological function, for example correct development or responses to environmental change. While this functional output can be modelled with intuition on participating molecules and mechanism \cite{bintu, ScholesSanchez}, an alternative approach can be to posit this function explicitly as an optimization goal and to find mechanisms consistent with this optimization \cite{Tkacik_2008, Baueretal, tkavcik2025information, sokolowski2025, mijatovic2025}. Optimization approaches have been of interest across biological sub-disciplines \cite{bialekbook}, starting with work on neurons \cite{efficientcoding, Laughlin}, as well as in the context of neural networks, where the loss function or optimization goal is perhaps more but also not always obvious.

Optimization in high dimensions can be difficult, but the high dimensionality of parameters and processes in biological systems provides an advantage to optimization approaches: optima of the loss-functions are often `sloppy’ \cite{gutenkunst}, which implies that along specific directions away from the optimum, combinations of parameters may vary significantly without affecting the loss function.  Therefore, for sloppy land-scapes even optimizing a mean performance will bring a solution `close’ to an optimum \cite{opt+var}. That multiple parameter combinations are consistent with near-optimal performance is indeed clear from data-driven theoretical observations – both for optimal regulatory elements \cite{Baueretal}, as well as for optimal signals \cite{Witteveen2025}- and also directly from experiment: in gene regulation, this observation is consistent with the biological diversity of functional regulatory pathways and constructs.  

Even though a challenge of optimization approaches is sometimes the absence of a clear, unifying goal (see e.g. the discussion in section V), optimization ideas can present a respectful approach for understanding how biological systems - in gene regulation and beyond -  fulfill their function. Nevertheless, it is important to critically assess the applicability of optimization goals, and the importance of constraints, in specific systems. Since living systems develop and evolve, investigating optimality also requires to take into account the kinetics or dynamic trajectories of the signal response,  an active area of research at the moment \cite{selimkhanov2014accurate, ahamed2021capturing, moor, reinhardt2023path, reinhardt2025mutual,  raju+siggia2023, gilpin2024generative, yadav2025homeorhetic}. An exciting general direction is to compare the regulatory and developmental dynamics to those of learning in biological and artificial systems, where optimization is more established and were dynamics can help predict outcomes, for example on whether networks learn or generalize \cite{nguyen2024differential}.

\section{Cells and Development}
 \subsection{Krishna Shrinivas}

The living cell is intrinsically high-dimensional. Vast numbers of microscopic parameters, such as molecular concentrations, interactions, and reactions, often tuned by evolution, drive emergent properties of low-dimensional cellular processes. A striking example of this funneling is the organization of the cellular milieu into tens of coexisting compartments called condensates whose molecular compositions comprise shared and distinct features. In recent years, phase transitions have been identified as central to driving condensate formation. Here I highlight few considerations that arise from viewing the cell as a squishy molecular stew rather than a watery, well-mixed soup of molecules. 

How do high-dimensional molecular properties channel into emergent condensate features, and which microscopic details—protein sequence, stoichiometry, excluded volume, reactions—actually matter? A related question is how best to encode such microscopic details and what could limit expressivity and capacity in resulting models \cite{zentner2025informationprocessingdrivenmulticomponent}. In parallel, biological measurements of condensates are typically low-dimensional - visualizable fluorophores on a handful of proteins, or phenotypic readouts such as expression of a downstream gene. From such sparse observables, can we infer higher-dimensional constraints on condensate properties and reveal evolutionary or functional restrictions on these ensembles? Statistical physics suggests possible routes: predictive models that coarse-grain microscopic degrees of freedom, response theories that infer hidden structure from perturbations, and information-geometric approaches that separate stiff from sloppy modes. These considerations resonate with other fields where distinct microscopic details nonetheless produce rich emergent behavior, offering opportunities for both importing methods as well as identifying unifying principles. Prominent examples discussed at the workshop include microbial ecology, where community-scale functions emerge from diverse consortia with distinct microscopic interactions, and understanding emergence in artificial intelligence, where billions of nodes are distilled into architectures that generalize and solve specific tasks. Ultimately, understanding the ingredients and recipes that underpin cellular stews offers a rich theoretical playground for high-dimensional biology. 

\subsection{Mohit Kumar Jolly}
Elucidating the dynamics of high-dimensionality of the regulatory biochemical space at multiple levels – transcriptional, epigenetic and metabolic – is an open question to understand cell decision-making in development, disease and reprogramming \cite{Rukhlenko2022-yl}. We have witnessed a surge in technological advancements that facilitate this information being collated in a dynamic manner experimentally and to the development of new computational methods that can extract at least pseudotime-based projections from static high-dimensional experimental data. Thanks to these advancements; we are inching closer towards mapping the changes in these high-dimensional landscapes that enable or accompany cell-state transitions, thus opening up new avenues to apply many dynamical systems theory approaches to identify which trajectories are allowed or not for cells traverse through this landscape, and how can their trajectories be modified to alter the destined cell-state \cite{Zhou2021-qc}. The concepts of attractor states, frustration and consequent cell-state canalization are being increasingly used to quantify entropy in gene expression during cell-state transitions, which can eventually allow for tying together the high-dimensional experimental data with low dimensionality of the phenotypic space, using theoretical biophysics-based novel conceptual frameworks \cite{JianhuaPRXLife, Hari2025-yw, Qiu2022-hn, Islam2025-qw}.

\subsection{Sanjay Jain}
Examples of biological systems that exhibit homeostasis or chaos both exist, but do we understand which systems exhibit these two behaviours, and when, and why? An example of homeostasis is the exponential growth culture of a bacterium, where hundreds to thousands of reactions involving similar numbers of molecular species achieve a coordinated steady state within each cell. This non-equilibrium steady state can be understood as a fixed point attractor of a complex dynamical system in a high-dimensional chemical concentration space  representing the cell in a fixed environment. Mathematical models of a cell involving the nonlinear chemical reaction kinetics of hundreds of interacting molecular species can be constructed that exhibit this behaviour \cite{samal2008regulatory,pandey2020exponential}. Stochasticity in the dynamics leads to a stable stationary probability distribution of states localized around the fixed point. This orderly behaviour, or homeostasis, is required for self-reproduction of the bacterium.
 
On the other hand, in this meeting we also saw examples of chaotic dynamics in biological systems from Greg Stephens (motion of C. Elegans \cite{ahamed2021capturing}, social behaviour of Zebrafish and honey bees). This prompts the question: What is the relationship between the homeostasis necessary for life and the observed chaotic behaviour? Can there be a classification of biological systems into those that exhibit stable fixed point or limit cycle type behaviour that could be regarded as homeostatic, and those that exhibit chaos? Are there common architectural/circuitry features (feedback, control or lack of it) that characterize each kind of behaviour? Is metabolism homeostatic and social behaviour chaotic? Are eukaryotes (and multicellular organisms) more prone to exhibiting chaotic dynamics compared to prokaryotes (and unicellular organisms)? Is chaotic behaviour always built on top of a homeostatic core? Do the same organisms show both homeostasis and chaos in different subsystems or different environments? Do these behaviours coexist (criticality or “edge of chaos” hypothesis)? Do the hundreds to thousands of interacting chemicals in {\it E. coli} have a dynamical regime which is chaotic? Maybe it is useful for the bacterium to be homeostatic while the going is good (food available, exponential growth). But when times are not so good (no food, stationary phase), it could be an advantage for the bacterium to be chaotic (more exploratory). It might be interesting to explore this in mathematical models and data.

\subsection{Sid Goyal}
Does the effective dimensionality of a biological system change with time or
conditions?
Our recent dynamical system view of hematopoiesis claimed that cells in the gene expression space display statistical signatures consistent with bifurcations during fate change and that there is a low-dimensional phase plane in gene expression space within which the multistability unfolds \cite{freedman2023dynamical}. Is this reduced dimensionality only evident during fate transitions? To address this, we asked if we can embed complete cell fate trajectories in low dimensions \cite{farrell2023inferring}  in a variety of development contexts. In doing so, we inferred a latent regulatory state that controls the dynamics of an individual cell to model multiple lineages. We find that the dimensionality of the
low dimensional space depended on the number of final states, suggesting some limitations on effective dimensionality. 

Is this low dimensionality a property developing systems getting canalized to provide robustness against perturbations? 
The expectation is that in contrast to the {\it in vivo} systems discussed above,  differentiated cells as they reprogram {\it in vitro}, can in principle take many
different paths in the high dimensional space of gene expression. However, we find only two populations
with distinct rates of reprogramming \cite{shakiba2019cell}. It will be fascinating if these correspond to two different
low dimensional manifolds, or is it an ensemble of different paths that give similar 
transition rates.

As we accumulate more examples of fate transitions and their dimensionality, it will be fascinating to understand if and how effective dimensionality is controlled and used in biology.

\subsection{Aneta Koseska}
Cells face a fundamental computational problem: they must process time-varying inputs utilizing signaling and genetic networks to make context-dependent decisions and determine cell fate \cite{Pertz2015, Witteveen2025, Rosen2025}. Understanding how the distributed biochemical networks accomplish this requires analyzing the evolving state space of the system, which is inherently high-dimensional. Non-autonomous dynamical systems provide a natural framework to describe the time-dependent trajectories of the network \cite{NonautoBook}. Our recent work demonstrates that biochemical networks in single cells perform a mapping of classes of dynamic input signals to classes of signaling state-space trajectories, each corresponding to a distinct phenotype \cite{yadav2025homeorhetic}. Crucially, this mapping requires the network to possess a form of dynamic memory \cite{Nandan_2022}, which is both necessary and sufficient to reproduce the observed behavior. Even low-dimensional coarse-grained models are capable of capturing this principle, highlighting the minimal mechanistic rules that govern how cells integrate complex inputs and commit to specific fates. At the same time, the high-dimensionality of the biochemical network enables robust separation of signaling trajectories, ensuring a unique mapping to each phenotype. This points to a subtle interplay between the rich structure provided by high-dimensional networks and the minimal features that low-dimensional models capture. Future work should focus on developing principled approaches to uncover how best to extract these minimal representations while preserving the functional benefits of high-dimensional biochemical architectures, providing a unified framework for understanding cellular computation in complex biological systems.

\subsection{William Gilpin}
Modern machine learning models exhibit the surprising ability to perform unseen logical tasks, like mathematical proofs, despite having never encountered them during training. This reasoning capability only emerges once models reach a sufficient number of parameters, implying that scale alone enables generalization capability. In such settings, dimensionality is an asset, not a curse. Why wouldn’t the same be true for biology? The brain distributes low-dimensional computations across thousands of neurons, while developmental pathways encode differentiation across hundreds of interacting genes. As in statistical learning, do these high-dimensional representations confer robustness, allowing biological systems to easily adapt to new tasks?  Emerging ideas from mechanistic interpretability in large-scale foundation models, like induction heads or attention routing, may thus have informative analogues in high-dimensional biological systems.

\subsection{Archishman Raju}
\begin{figure}
\begin{center}
\includegraphics[width=\linewidth]{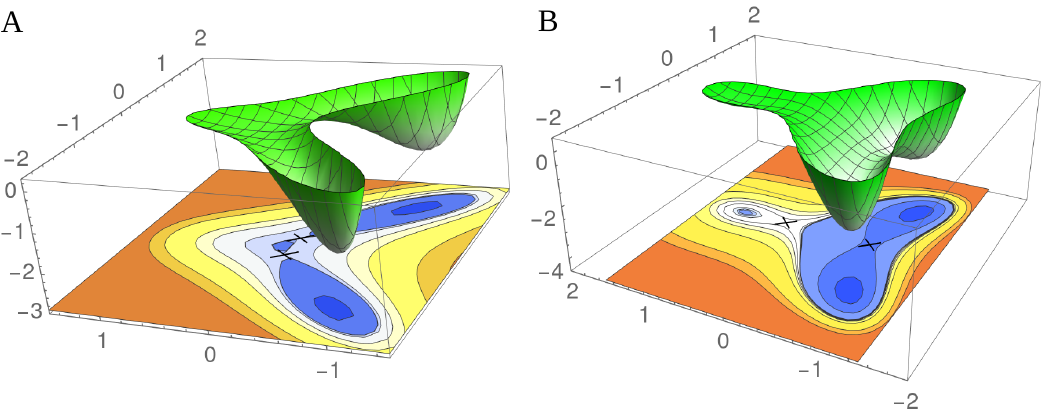}
\caption{{\bf Waddington landscapes constitute a low dimensional representation of cell fate decisions.} The figures shows two distinct potential landscapes that could putatively describe the differentiation of a cell into two alternative types. These constitute different \textit{geometrical} possibilities for the low dimensional behaviour. \textbf{A)} There is a direct path from the central attractor into the two alternative attractors.  \textbf{B)} On exiting the central attractor, the flow goes towards the second saddle point and is subsequently diverted towards one of the attractors. Adapted from Ref. \cite{raju2024geometrical}.
}
\label{fig:Waddingtonlandscape}
\end{center}
\end{figure}

Organismal Development is shaped by processes that involve complex gene regulatory networks and signaling pathways which have been characterized in great detail in model organisms over the past several decades. However, our ability to quantitatively understand high dimensional gene expression data remains limited and we continue to rely on heuristic computational techniques to understand this data. This difficulty was, in many ways, anticipated by the work of a section of early embryologists and theoretical biologists who established a connection between development and dynamical systems theory and were pessimistic about the usefulness of a reductionist approach to developmental phenomena. Dynamical systems theory has a long tradition of first qualitatively characterizing the behaviour of a system of differential equations and then quantitatively understanding the simplest ``normal form" equations consistent with the qualitative behaviour, which may characterize the universal behaviour of a multiplicity of biological mechanisms. These mathematical insights need to be converted into calculational tools that can be directly connected to biological data. Recent attempts to formalize Waddington landscapes (Fig.~\ref{fig:Waddingtonlandscape}) in the context of cell fate specification fall into this broader project~\cite{rand2021geometry, saez2022statistically, raju+siggia2023, raju2024geometrical}. It is possible that these ideas are applicable in other contexts particularly in cases where an underlying high dimensional system has low dimensional outcomes. 

\subsection{Shaon Chakrabarti}
What constitutes `good' low dimensional representations of complex datasets and how to achieve such latent space embeddings, are interesting open questions \cite{bengio_representation_2013}. Recent advances in generative AI such as Variational Autoencoders (VAEs) \cite{kingma_auto-encoding_2014} and Generative Adversarial Networks (GANs) \cite{goodfellow_generative_2014} provide interesting solutions to these questions in the context of identifying biological cell-states, by disentangling multiple sources of variation in low dimensional spaces \cite{lopez_deep_2018,marouf_realistic_2020}. However, it remains unclear how functionally relevant these inferred cell-states are, especially when the ground truth is unknown (as is typically the case). It also remains unclear how biological and measurement noise affects the cellular resolution achievable in the inference of these cell-state embeddings. While cell-states emerge from many biological processes working together, current measurement technologies can only access a small fraction of these processes simultaneously. Random fluctuations in these few measured modalities may therefore blur the boundaries between cell-states and set fundamental limits to their identification. 

For example, such a limitation has recently been demonstrated in the specific context of the mammalian circadian clock, where cells in different clock phases correspond to biologically relevant (ground truth) cell-states \cite{nikhat_transcriptional_2025}. While population averaged RNA levels can be used to learn these correct cell-states on a circular 2D latent space, single-cell RNA measurements are too noisy and generate incorrect latent spaces. This is true even if technical noise associated with the measurement technique is minimized \cite{nikhat_transcriptional_2025}. This therefore serves as an example of how biological noise can affect the information content of low-dimensional embeddings generated by deep learning algorithms. Here, approaches from Information Theory could provide deeper insight into the performance of these algorithms, by quantifying how much of the information content of the high dimensional cell-states is captured in the latent space. Information Theory could also be leveraged to predict how much further increase in the latent space information content might be theoretically possible, that could guide improvements in the deep neural network architecture and loss functions. 

\section{Ecology, Evolution and Immunology}
\subsection{Michael Desai}
Evolution is a process by which biology explores high-dimensional genotype-phenotype landscapes through random mutations, stochastic genetic drift, and natural selection. It might seem impossible that this process could reliably lead to adaptation in the face of diverse and ever-changing biological challenges. Yet evolution can solve many of these problems routinely and repeatably. How and why does this work? It may be that, instead of representing an obstacle, the high dimensionality of genotype and phenotype spaces leads to predictable emergent regularities that evolution can reliably exploit. Further, evolution may have led to regions of these spaces which have properties that are amenable to further adaptation. There may be precise connections that one can draw between these questions and the training of large machine learning models, and there is the potential for practical lessons from evolution to influence machine learning and vice versa. In particular, questions about the emergence and consequences of various forms of modularity were highlighted in the workshop and are clearly broadly important across these fields. 

\subsection{Jacopo Grilli}
High-dimensionality is both the central challenge and the defining feature of community ecology—how do hundreds of species coexist despite complex interactions? A crucial distinction is that unlike neural or cellular systems where optimality principles (minimizing energy, maximizing information processing) provide powerful organizing frameworks, ecological communities lack any obvious optimization target—there is no fitness function for a community, only for individuals within it. This raises fundamental questions: under what conditions can individuals actually behave optimally in high-dimensional environments where the fitness landscape constantly shifts due to other species? What even defines optimality is context-dependent (growth rate, yield, robustness), and the mismatch between individual-level optimization and community-level patterns is itself a source of emergent complexity. The challenge is compounded by dynamical feedback between organisms and environment, where both are high-dimensional: organisms possess many traits and the environment comprises many factors, creating a complex co-evolutionary landscape where causes and effects are deeply entangled. Techniques from statistical physics and random matrix theory—scaling arguments, disorder averaging, studying typical rather than specific cases—offer ways forward by extracting universal patterns (macroecological laws, stability conditions) that emerge precisely because high-dimensionality washes out microscopic details. Conversely, ecology exports this crucial insight to other high-dimensional systems: the interplay between individual optimization, environmental feedback, and emergent collective behavior may be more fundamental than seeking system-level optimality principles, with applications from immune repertoires to artificial ecosystems where no designer's objective exists.

\begin{figure*}[hbt!]
\begin{center}
\includegraphics{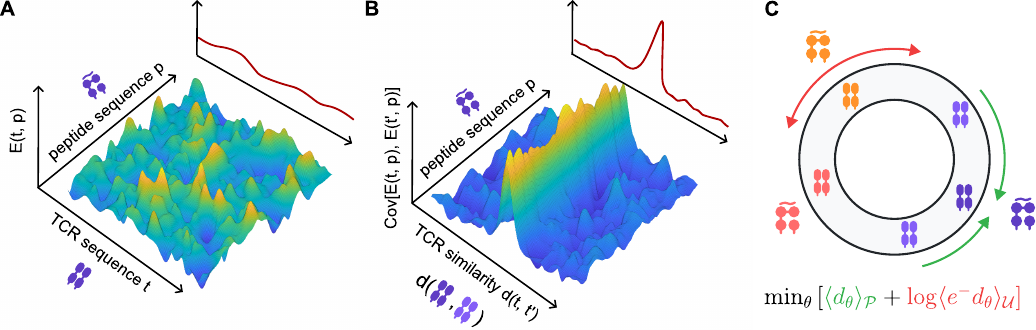}
\caption{{\bf Contrastive learning can reveal generalizable similarity rules in the complex sequence to function map of immune receptors and their ligands.} Cartoons of \textbf{A)} a binding energy landscape describing T cell receptor specificity to different peptide ligands and \textbf{B)} the co-variance of this landscape with receptor sequence similarity. By 'aligning' the peaks in the landscape the co-variance structure is expected to generalize to a greater extent across ligands as illustrated by the marginal distributions (insets). \textbf{C)} Contrastive learning optimizes receptor similarity metrics to minimize the distance between receptors which are co-specific to a common ligand, while encouraging uniform use of the representation space.  [Adapted from Ref. \cite{pyo2025data}].
}
\label{fig:ContrastiveLearningTCRs}
\end{center}
\end{figure*}

\subsection{Akshit Goyal}
There is a longstanding tradition in physics of replacing complexity with randomness when it comes to complex, high-dimensional systems. This approach traces back to Wigner and Dyson’s seminal work in the 50s, who used random matrix theory to successfully predict statistical properties of the uranium nucleus \cite{wigner1951statistical,dyson1962statistical}. The key insight has always been to use randomness while respecting the fundamental constraints of the system of interest; for Wigner, this meant preserving the symmetry of the random Hamiltonian. May adopted this philosophy remarkably early in ecology (1972), modeling interspecies interactions as random matrices to probe the stability of complex, species-rich ecosystems \cite{may1972will}. Because ecological dynamics operate with a large number of species, the random matrix approach offered a tractable route to navigate high-dimensional complexity. Over the past decade, this idea has had a resurgence, resulting in new predictions about many statistical properties of ecosystems \cite{nemenman2025randomness,cui2024houches,bunin2017ecological}.

Yet a persistent criticism remains: these approaches use the ``wrong'' random ansatz, with the familiar refrain from biologists that ``ecosystems are not random''. While everyone agrees that species and their interactions possess structure, determining what this structure should generically look like without imposing it by hand has been challenging. This concern extends across biological systems, e.g., intracellular condensates, metabolic networks, and other contexts where randomness has been invoked. 

An emerging resolution is that for biological systems, the appropriate constraints may arise from evolution itself. In preliminary work on eco-evolutionary dynamics, we find that even when communities begin with random, unstructured interactions, continual feedback between ecological and evolutionary processes produces communities that remain diverse yet develop highly structured interaction patterns \cite{feng2025theory}. Surviving strains become strongly phenotypically correlated and the evolved interaction structure is consistent and predictable. This work gives us hope that evolutionary stability is a missing ingredient that we must account for when thinking about ecosystems. The implication is potentially unifying: disordered interactions with evolutionarily stable structure may provide the ``right'' ansatz for studying high-dimensional biological systems.

\subsection{Andreas Tiffeau-Mayer}
Immunology is a paragon of high-dimensional complexity in biology. The trillion lymphocytes in our bodies carry more than a hundred million different antigen receptors drawn at random from an even more diverse set of possibilities. Yet, luckily most of us are typically able to mount an immune responses to any novel pathogen challenge. This crucially hinges on the multitude of receptors that bind any given target. Recent progress in deciphering the many-to-many mapping between immune receptors and pathogens has begun to reveal the biophysical rules underlying this redundancy in the coverage of antigen space.

As in other areas of protein biophysics, contrastive learning has emerged as a key strategy for learning function-aware coarse-graining procedures (Fig.~\ref{fig:ContrastiveLearningTCRs}). In the immunological context, this approach has uncovered statistical rules about which receptors are likely to bind the same ligand, revealing for example the importance of steric constraints in these interactions \cite{pyo2025data}. Importantly, knowledge of such rules enable detecting responses in immune repertoires by identifying convergently selected clusters of receptors with related binding specificity. Taming the immense receptor diversity by such coarse-graining approaches has thus emerged as a key advance in the study of the adaptive immune system, with applications in precision immunomonitoring and vaccine design.

Yet, if the aim is to develop a global theory of immune repertoire ecology, the description of immunological diversity at the level of these clusters of cells remains too high-dimensional for identifiability of parameters from current experimental data. Therefore techniques  replacing detailed assumptions about interactions between immune cells and pathogens with random ensembles of fluctuating forces also play an important role in immunology \cite{desponds2016fluctuating, gaimann2020early}. A potential future direction lies in the application of techniques developed for the analysis of longitudinal microbiome dynamics to reveal 'guild structure', pinpointing co-regulation of immune cells at scales beyond shared antigen specificity.

\section{Behaviour, Learning and Larger Scales}

\subsection{Greg Stephens}
In the physical study of animal behavior, dimensionality has repeatedly emerged as a key organizing principle. Across diverse organisms, the space of body postures sampled from freely-moving conditions is low-dimensional, sometimes dramatically so, especially when compared to the number of underlying components such as neurons or muscles from which behavior emerges.  A classic example is the nematode worm {\em C. elegans}, where four ``eigenworm'' directions dominate the shape covariance matrix \cite{stephens2008}, even for a wide variety of mutants \cite{Brown2013}.

However, a more contentious and contemporary issue is the dimensionality of behavioral dynamics—how body configurations change over time. For example, recent analyses of {\em C. elegans} foraging have shown that worm behavior can be effectively described as a low-dimensional (seven!) chaotic dynamical system \cite{ahamed2021capturing}. Further, the complexity of this behavioral attractor can be manipulated; for example six additional dimensions emerge from behavior in environments seeded with {\em E. coli} but these can then be eliminated with an appropriate mutation.

In this view, complex posture dynamics and multiple timescales arise from low-dimensional but nonlinear interactions, which is reminiscent of the onset of turbulence \cite{Brandstater1983}.  In contrast, long-term recordings of walking flies reveal an apparent power-law dynamics \cite{bialek2024long}. Such power-law dynamics are at odds with a Markovian or low-dimensional framework, and one proposed resolution is that they result from time-varying internal states \cite{costa2024fluctuating}.

Another intriguing possibility however, is that these findings in behavior mimic our early (mis)understanding of critical phenomena. For example, in the 2D Ising model the mean-field magnetization undergoes an order-disorder transition, but with the wrong critical exponents - this is fixed only by considering the presence of (high-dimensional) fluctuations. Interestingly,
this dimensional dichotomy is mirrored in neuroscience: while some studies point to low-dimensional neural manifolds (see e.g.~\cite{perich2025neural}), others reveal power-law activity \cite{manley2024simultaneous}. Resolving the tension between low-dimensional attractors and power-law behavior has the potential to reveal fundamental principles of both behavior and neural computation (see e.g.~\cite{fontenele2024low}).

\begin{figure}[hbt!]
\begin{center}
\includegraphics[scale=0.25]{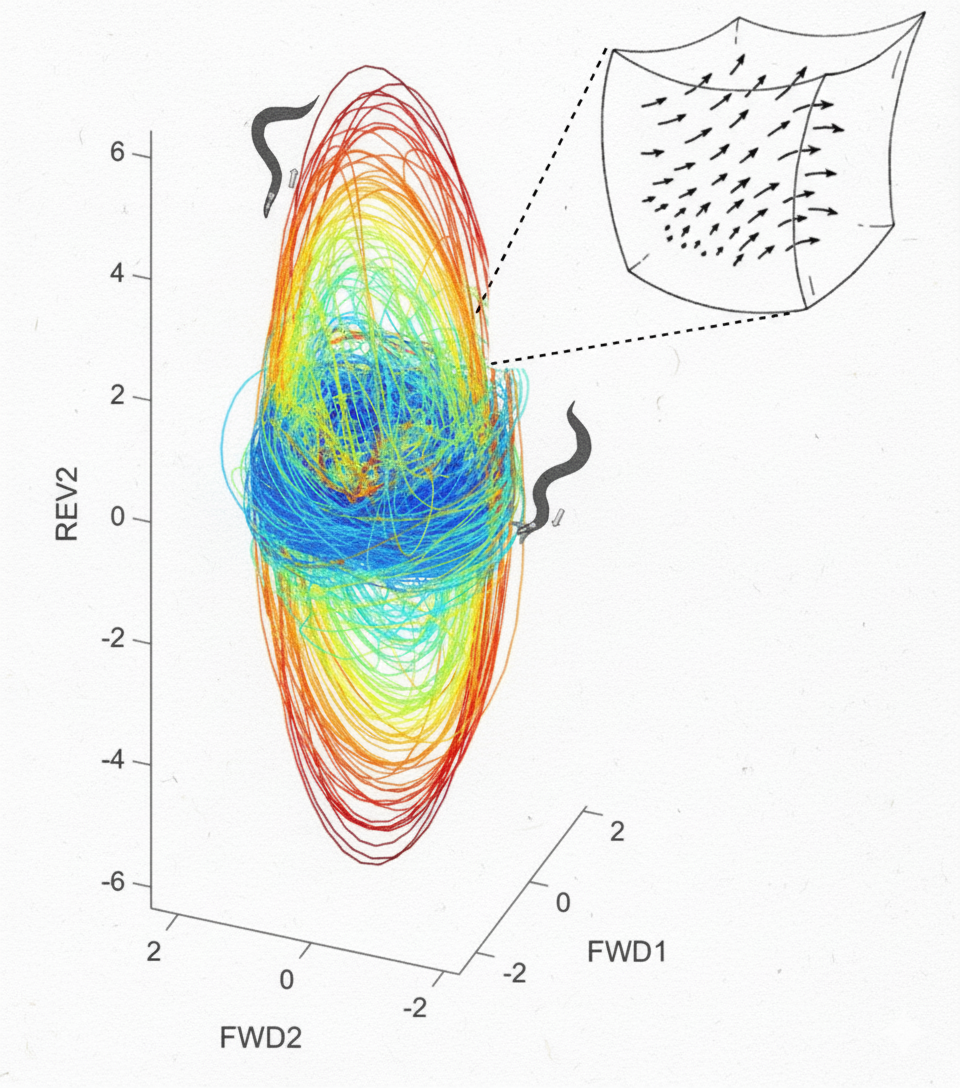}
\caption{{\bf Low-dimensional dynamics in the behavior of {\em C. elegans}}. We show a 3D projection of embedded worm posture trajectories illustrating forward (blue) and backward (red) locomotion through approximately orthogonal collections of cycles, a simplification of the full 7D state space \cite{ahamed2021capturing}. Cartoon arrows are a reminder that the geometry of this space carries fundamental dynamical information, such as the Lyapunov exponents. 
Figure credit: Tosif Ahamed.}
\label{fig:StateSpaceSketch}
\end{center}
\end{figure}

\subsection{Simone Pigolotti}
Throughout its history, physics has achieved remarkable success in describing a wide range of phenomena with elegant, simple, and accurate mathematical models. Notable examples include Newton's laws, Maxwell's equations, and the standard model of particle physics. Despite these successes, biology presents a formidable challenge. Even the simplest cells contain thousands of genes with different functions. This daunting complexity makes it unclear whether theory in biology will ever be as successful as in physics. 

Two recent developments have dramatically changed the game. First, modern experimental techniques provide access to an unprecedented wealth of high-dimensional, quantitative data on biological systems. Second, the AI revolution has granted us tools to extract patterns from these data, and reduce these high-dimensional datasets to relatively few relevant variables. 

These innovations offer a rather direct approach to our problem: gather as much high-quality data as possible and let AI discover the ``model'' for us. However, the reduced descriptions found by AI are not always easy to interpret. Moreover, not all modern biological data are high-throughput and accurate: in many areas, we are still dealing with limited and noisy datasets. 

Achieving a deep understanding of biology will likely require a clever combination of traditional theory and data-driven AI methods. How to best integrate these two approaches remains an exciting open question.

\subsection{Gautam Reddy}
Much of the complexity of behavior arises from the fact that an animal’s decision to make a move or take a path is fundamentally intertwined with the statistics of the environment it interacts with. Survival requires informed decisions, and informed decisions require an animal to store a representation of its history of movements and observations. But movements and observations are often ``high-dimensional,'' and the number of possible histories is exceedingly large. Fortunately, most information is not useful for behavior, and it is reasonable to believe that animals form representations that are useful specifically for guiding future, rewarding behavior rather than reconstructing past histories.

Representations of an animal’s history and its own drives (such as hunger) form a persistent internal state. What is the “dimensionality” of this internal state space? How are internal states represented in neural systems? And how do they modulate context-dependent behavior? The capacity to precisely measure behavior, neural activity, connectomics, single-cell transcriptional states, and other modalities of neural communication offers hope that some of these questions will be resolved in the near future.

In parallel, recent developments in large language models (LLMs) offer an intriguing alternative path. LLMs do indeed develop causally identifiable internal states, and this capacity “emerges” after training large models on large amounts of data to predict what comes next, that is, without explicit human intervention. The computational substrate on which LLMs operate is, of course, different from biology, and the mechanistic details of how an internal state may arise will certainly be different. But it is natural to ask whether one can draw broader conceptual insights into the nature of internal states from a rigorous empirical study of LLM behavior. 

\section{Conclusion}
As the datasets that describe the biological systems increase both in quality and dimensionality, several theoretical approaches towards understanding seem to emerge. One set of approaches targets the identification of an effectively low dimensional description, which could, for some systems, justify and lead to the development of simple predictive microscopic models. Practically, a low-dimensional description can be found by compression of the larger space, or the identification of a representative latent space. This low dimensional structure could also be described with a sloppy or soft mode framework. An interesting question is how such an effective low-dimensionality is present across scales. Hypotheses that may explain this include stereotyped evolutionary or developmental trajectories that only allow the system to explore a limited part of state space. Alternatively, internal feedbacks or external constraints and performance optimizations could have driven the system towards specific subspaces. 

Another set of approaches towards a unifying theory lies in embracing the high dimensionality: for example, by exploiting access to the statistics to fit a description or train a model. This high-dimensionality can apply both to a particular state as well as to the dynamics. Dynamic trajectories can retain high-dimensionality and partially chaotic behaviour, which can nevertheless be quantified; alternatively, they may remain high-dimensional because they track an externally changing landscape. In this context, the dimensionality can be an asset to both biological and computational systems that allows them to fulfill multiple tasks using a given set variables.

Future directions of research could involve an improved quantification of the effective dimensionality in the system and its dynamics, identifying drivers that lead to the emergence of specific structured sub-spaces in a specific biological question, and develop analogies between biological and artificial systems across scales.

\section{Acknowledgements}
This research was supported in part by the International Centre for Theoretical Sciences (ICTS) for participating in the program - Unifying Theories in High-Dimensional Biophysics (code: ICTS/UHDB2025/07), as well as the American Physical Society DBIO (Division of Biological Physics).


\bibliography{ref_uhdb}
\end{document}